\documentclass[aps,prr,10pt,showpacs,amsmath,amssymb,twocolumn,nobibnotes,preprintnumbers,longbibliography]{revtex4-2}
\usepackage[colorlinks=true,citecolor=blue,linkcolor=red,urlcolor=red]{hyperref}
\usepackage[sort&compress]{natbib}
\usepackage{graphicx}
\usepackage{amssymb}
\usepackage{braket}
\usepackage{amsmath}
\usepackage{hyperref}
\usepackage{dsfont}
\usepackage{bm} 
\usepackage{physics}
\usepackage{tabularx}
\usepackage{verbatim}
\usepackage{float}
\usepackage{romannum}
\usepackage{bbold}
\usepackage{soul}
\usepackage[normalem]{ulem}
\usepackage{aligned-overset}
\usepackage{multirow}
\usepackage{tabularx}
\usepackage{cancel}
\usepackage{relsize}

\renewcommand\r{{\bf r}}
\newcommand\p{{\bf p}}

\newcommand\x{{\bf x}}

\begin{document}
	
\author{Anirudh Gundhi}
\email{anirudh.gundhi@units.it}
\affiliation{Department of Physics, University of Trieste, Strada Costiera 11, 34151 Trieste, Italy}
\affiliation{Istituto
	Nazionale di Fisica Nucleare, Trieste Section, Via Valerio 2, 34127 Trieste,
	Italy}

\author{Oliviero Angeli}
\email{oliviero.angeli@phd.units.it}
\affiliation{Department of Physics, University of Trieste, Strada Costiera 11, 34151 Trieste, Italy}
\affiliation{Istituto Nazionale di Fisica Nucleare, Trieste Section, Via Valerio 2, 34127 Trieste, Italy}

\author{Angelo Bassi}
\email{abassi@units.it}
\affiliation{Department of Physics, University of Trieste, Strada Costiera 11, 34151 Trieste, Italy}
\affiliation{Istituto
	Nazionale di Fisica Nucleare, Trieste Section, Via Valerio 2, 34127 Trieste,
	Italy}
	
\title{From equivalent Lagrangians to inequivalent  open quantum system dynamics}
	
\date{\today}

\begin{abstract}
Lagrangians can differ by a total derivative without altering the equations of motion, thus encoding the same physics. This is true both classically and quantum mechanically. We show, however, that in the context of open quantum systems, two Lagrangians that differ by a total derivative can lead to inequivalent reduced dynamics. While these Lagrangians are connected via unitary transformations at the level of the global system-plus-environment description, the equivalence breaks down after tracing out the environment. We argue that only those Lagrangians for which the canonical and mechanical momenta of the system coincide lead to operationally meaningful dynamics. Applying this insight to quantum electrodynamics (QED), we derive the master equation for bremsstrahlung due to an accelerated non-relativistic electron upto second order in the interaction. The resulting reduced dynamics predicts decoherence in the position basis and closely matches the Caldeira–Leggett form, thus resolving previous discrepancies in the literature. Our findings have implications for both QED and gravitational decoherence, where similar ambiguities arise.
\end{abstract}

\maketitle
\section{Introduction}
\label{Intro}
Decoherence in quantum systems \cite{KieferDecoherence, Schlosshauer2007,Joos:1984uk, zeh2002decoherence} is ubiquitous and unavoidable. It is relevant in a large variety of physical situations such as matter-wave interferometry \cite{Brezger2002,Hornberger2003,Lucia2003,hackermuller2004decoherence,Cotter2015,Cotter2017}, quantum optics \cite{Bullier_2020,Gonzalez-Ballestero2021, Gosling2024}  and quantum computers \cite{Google2023, kim2023evidence}, to name a few. Over the years, several theoretical models~\cite{WIGNER1984, Joos:1984uk, GandF, BassanoCollisions, CALDEIRA1983, Caldeira1985, UnruhZurek1989, HuPazZang1992, Tegmark:1993yn} have been developed to understand how a system behaves effectively under the influence of its environment, and how this influence can be shielded or at least mitigated, or perhaps exploited for specific tasks. 

A natural question concerns the derivation of these models, most of which are phenomenological, from the fundamental Lagrangian describing the interaction between the system and its environment. Given that gravity is weak and nuclear forces are irrelevant for the typical systems of interest, this Lagrangian reduces to that of Quantum Electrodynamics (QED). The derivation of these models is not straightforward---even at the non-relativistic level in which we will operate.

Physical intuition and previous theoretical research suggest that in standard conditions the reduced dynamics should resemble the Caldeira-Leggett master equation  \cite{CALDEIRA1983,KieferDecoherence}:
\begin{align}\label{eq:cl}
\dot{\hat{\rho}}_r = -\frac{i\eta}{2m}\left[\hat{x},\{\hat{p},\hat{\rho}_r\}\right]-\frac{\Lambda}{\hbar}\left[\hat{x},\left[\hat{x},\hat{\rho}_r\right]\right]\,,
\end{align}
where the coefficients $\eta$ and $\Lambda$ should be derived as functions of the parameters of the Lagrangian. The Caldeira-Leggett equation models a two-way exchange of energy between the system and environment. In particular, the first term on the right describes energy lost by the system to the environment, and the second term describes decoherence of the system in the position basis due to \textit{measurements} made on the system by the environment.

However, the literature presents contrasting results in this regard.
For example, the case of a quantum particle subject to vacuum  fluctuations of the electromagnetic (EM) field has been studied in \cite{Caldeira1991, AnastopoulosQED, BandP,Gundhi:2023vjs} and most recently in \cite{Gundhi_MeasDec_2025}. (See also \cite{Kiefer:1992, Ford, Baym_Ozawa, Santos1994, Diosi1995} for a related discussion.) However, while \cite{Caldeira1991,Gundhi:2023vjs,Gundhi_MeasDec_2025} arrive at results compatible with Eq.~\eqref{eq:cl}, predicting in particular decoherence in position basis, others \cite{AnastopoulosQED, BandP} arrive at substantially different conclusions. The source of this discrepancy plagues not only QED but also the recent works concerning gravitational decoherence \cite{Anastopoulos1996, Anastopoulos_2013, Blencowe2013, Oniga2016, Kanno2021, Lorenzo2021, Lagouvardos2021, Wilczek2021, Carney2023}, a subject that has attracted a lot of attention as an indirect probe for the quantum nature of gravity. 

The goal of this work is to demonstrate that this issue arises whenever one attempts a microscopic derivation of the reduced dynamics starting from a Lagrangian.  
Specifically, we will show that the root of the problem lies in the choice between two Lagrangians that differ by a total derivative. This difference is unimportant at the level of equations of motion derived classically, or at the level of full expectation value of operators in quantum mechanics. This is because the operators and statevectors for the quantum dynamics derived from the two Lagrangians are related to each other by a unitary transformation \cite{Ackerhalt84,TannoudjiPartOneChapterFour}. 

However, this is not the case for open quantum systems since such a unitary transformation mixes the system and the environment, and the equivalence is lost upon tracing over the environment (which is a non-unitary operation), needed to arrive at the reduced dynamics for the system alone. Faced with this in-equivalence, one must then pick one Lagrangian over the other. We will show that from the operational point of view, one must start with a Lagrangian for which the canonical and the mechanical momentum of the system coincide.

In the light of these results, starting from  the appropriate form of the standard QED Lagrangian, we derive the master equation for an accelerated non-relativistic electron which decoheres due to which-path information carried away by bremsstrahlung. As expected, the master equation we derive resembles closely the phenomenological master equation~\eqref{eq:cl} which is typically used to describe decoherence. In particular, our master equation for bremsstrahlung predicts decoherence in the position basis, in contrast to decoherence in the conjugate momentum basis as in Ref.~\cite{BandP_Brem2001}. We thus recover established (phenomenological) results in the literature \cite{CALDEIRA1983, Caldeira1985}  providing further support in favor of our criterion.

As already mentioned, this analysis is relevant not only for electrodynamics, but also for (linearized) gravity, whose Lagrangian allows for a similar freedom in its definition, and therefore a similar ambiguity which has to be resolved. 

\section{In-equivalence between master equations: a case study}\label{sec:InEq} 

Both in classical and quantum physics, two Lagrangians that differ by a total derivative describe the same physical situation. We will show through a simple example that this equivalence does not hold for open quantum systems, where  by definition one has to distinguish between a system and its environment, and trace over the environment to obtain a master equation for the system alone.

Let us consider two non-relativistic particles moving in one dimension, where particle 1 is taken to be the system, and particle 2 is treated as its environment. Suppose the Lagrangian is: 
\begin{align}
L=\frac{1}{2}m_1 \dot{x}^2_1 + \frac{1}{2}m_2 \dot{x}^2_2+ gx_1\dot{x}_2,
\end{align}
or equivalently:  
\begin{align}
L'=\frac{1}{2}m_1 \dot{x}^2_1 + \frac{1}{2}m_2 \dot{x}^2_2-g\dot{x}_1x_2,
\end{align}
since the two differ by a total derivative:
\begin{align}\label{eq:Ftoy}
L=L'+\frac{d}{dt}(gx_1x_2).
\end{align}
We  quantize the system starting from these two Lagrangians, and then compute the reduced dynamics for particle 1, showing that they are substantially different.

{\it Open system dynamics from $L$.--}
Following standard canonical quantization, we first have to derive the conjugate momenta for $x_1$ and $x_2$; they are:
\begin{align}\label{eq:momenta1}
p_1 = \frac{\partial L}{\partial \dot{x}_1} = m_1 \dot{x}_1\,,\qquad p_2 = \frac{\partial L}{\partial \dot{x}_2} = m_2 \dot{x}_2+gx_1.
\end{align}

The Hamiltonian then reads:
\begin{align}\label{eq:H_L}
H = \frac{p^2_1}{2m_1}+\frac{g^2x^2_1}{2m_2}+\frac{p^2_2}{2m_2}-\frac{gx_1p_2}{m_2}.
\end{align}
Quantization proceeds as usual, with $(\hat x_1, \hat p_1)$ and $(\hat x_2, \hat p_2)$ as pairs of conjugate variables satisfying the canonical commutation relations.

The reduced density matrix for particle 1 is:
\begin{equation} \label{eq:rdm}
\hat \rho_1(t) \equiv \text{Tr}^{(2)}[\hat \rho(t)] = \int dx_2 \, \langle x_2 | \hat \rho(t) | x_2\rangle,
\end{equation}
where $\text{Tr}^{(2)}[\cdot]$ denotes the partial with respect to particle 2. Through standard calculations, presented in Appendix~\ref{appendix:B}, one arrives at the following master equation for $\hat \rho_1$, which is valid up to second order in the coupling constant $g$:
\begin{align}\label{eq:Master_L1}
\partial_t\hat{\rho}_1=&-\frac{i}{\hbar}\left[\frac{\hat{p}^2_1}{2m_1}+\frac{g^2\hat{x}^2_1}{2m_2},\hat{\rho}_1\right]\nonumber\\
&-\frac{g^2\langle\hat{p}^2_2\rangle_i}{\hbar^2m_2^2}\left(t[\hat{x}_1,[\hat{x}_1,\hat{\rho}_1]]-\frac{t^2}{2m_1}[\hat{x}_1,[\hat{p}_1,\hat{\rho}_1]]\right),
\end{align} 
with $\langle\hat{p}^2_2\rangle_i = \mathrm{Tr}\{\hat{\rho}_2(t_i)\hat{p}^2_2\}$.
This master equation predicts decoherence in the $x$ basis and involves the operator $\hat{p}_1$ (conjugate to the position operator $\hat{x}_1$), whose eigenvectors are the same as the eigenvectors of the velocity operator $\hat{v}_1$.

{\it Open system dynamics from $L'$.---}
We now repeat  the same procedure, starting from the Lagrangian $L'$.
The conjugate momenta for $x_1$ and $x_2$ for $L'$ are different compared to $L$, and are given by:
\begin{align}\label{eq:momenta2}
p'_1 = \frac{\partial L'}{\partial \dot{x}_1} = m_1 \dot{x}_1-gx_2,\qquad p'_2 = \frac{\partial L'}{\partial \dot{x}_2} = m_2 \dot{x}_2.
\end{align}
The Hamiltonian corresponding to $L'$, in terms of the canonical variables, reads:
\begin{align}\label{eq:MasterEquation2}
H' = \frac{p'^{2}_1}{2m_1}+\frac{p'^2_2}{2m_2}+\frac{g^2x^2_2}{2m_1}+\frac{gx_2p'_1}{m_1}.
\end{align}
Note that $H'$ coincides with $H$, once $p_{1}'$ and $p_{2}'$ are replaced with $p_{1}$ and $p_{2}$ via the relations $p_{1}' = p_1 - gx_2$ and $p_{2}' = p_2 - gx_1$.

Canonical quantization now requires that $(\hat x_1, \hat p_1')$ and $(\hat x_2, \hat p_2')$ are the pairs of conjugate variables satisfying the canonical commutation relations~\footnote{This is consistent with the previous choice of conjugate variables, since one set of commutation relations implies the other.}.

The reduced density matrix $\hat \rho_1'$ for particle 1 is again defined as in Eq.~\eqref{eq:rdm}, but now its master equation,  valid up to second order in the coupling constant $g$, is completely different:
\begin{align}\label{eq:Master_L2}
\partial_t{\hat{\rho}'}_1=&-\frac{i}{\hbar}\left[\frac{\hat{p}'^2_1}{2m_1}\left(1-\frac{g^2t^2}{2m_1m_2}\right),\hat{\rho}'_1\right]-\frac{g^2}{\hbar^2m_1^2}\left(t\langle\hat{x}^2_2\rangle_i\phantom{\frac{1}{2}} \right.\nonumber\\
&\left.+\frac{t^3\langle\hat{p}'^2_2\rangle_i}{2m^2_2}+\frac{3t^2}{4m_2}\langle\{\hat{x}_2,\hat{p}'_2\}\rangle_i\right)[\hat{p}'_1,[\hat{p}'_1,\hat{\rho}'_1]].
\end{align}
with $\langle\hat{x}^2_2\rangle_i = \mathrm{Tr}\{\hat{\rho}_2(t_i)\hat{x}^2_2\}$, $\langle\hat{p}'^2_2\rangle_i = \mathrm{Tr}\{\hat{\rho}_2(t_i)\hat{p}'^2_2\}$ and $\langle\{\hat{x}_2,\hat{p}'_2\}\rangle_i = \mathrm{Tr}\{\hat{\rho}_2(t_i)\{\hat{x}_2,\hat{p}'_2\}\}$. 
We immediately see that the two master equations~\eqref{eq:Master_L1} and~\eqref{eq:Master_L2} describe different time evolutions of particle 1. 

Contrary to Eq.~\eqref{eq:Master_L1}, the master equation~\eqref{eq:Master_L2} predicts decoherence in the basis of $\hat{p}'_1$. For the Lagrangian $L'$, $\hat{p}'_1$ is conjugate to $\hat{x}_1$ but is different from the velocity operator, i.e., $\hat{v}_1\neq\hat{p}'_1/m_1$. Furthermore, they yield different equations of motion  for the expectation value of the position operator   $\langle\hat{x}_1\rangle_t$, as shown in Appendix~\ref{appendix:B}. We shall now describe why this is the case.

\section{Why different master equations}\label{sec:equivalence} 
When two Lagrangians differ by the total derivative of a given  function $F(x_1,x_2)$, 
\begin{align}\label{eq:Lag}
L(x_1,\dot{x}_1,x_2,\dot{x}_2) = L'(x_1,\dot{x}_1,x_2,\dot{x}_2)-dF/dt,
\end{align}
the conjugate momenta for the two Lagrangian descriptions are different, and are related to each other as
\begin{align}\label{eq:ConjMom}
    p_{(1,2)} = p'_{(1,2)}-\frac{\partial F}{\partial x_{(1,2)}},
\end{align}
where $p_{(1,2)}$ stands for $p_1$ or $p_2$. In $L$, a given observable $\mathcal{O}$ is expressed as a function of $\big(\hat{x}_{(1,2)},\hat{p}_{(1,2)}\big)$, while in $L'$ it is expressed  as a function of $\big(\hat{x}_{(1,2)},\hat{p}'_{(1,2)}\big)$. Due to the difference between the conjugate momenta of the two descriptions, the functional form of $\mathcal{O}$ in $L$, which we denote by $\hat{O}(\hat{x}_{(1,2)},\hat{p}_{(1,2)})$, is different from its functional form in $L'$, which we denote by $\hat{O}'(\hat{x}_{(1,2)},\hat{p}'_{(1,2)})$. An example is how the Hamiltonian $\mathcal{H}$ takes different functional forms $H$ and $H'$ in Eqs.~\eqref{eq:H_L} and~\eqref{eq:MasterEquation2} respectively. 

Nevertheless, given the functional form of $\hat{O}$ in $L$, one can determine the functional form of $\hat{O}'$ in $L'$ (and vice-versa). This relationship is determined by $\hat{F}$ and is given by~\cite{Ackerhalt84,TannoudjiPartOneChapterFour}:
\begin{align}\label{eq:Unitary}
\hat{O'}\big(\hat{x}_{(1,2)},\hat{p}'_{(1,2)}\big) = \hat{T}\hat{O}\big(\hat{x}_{(1,2)},\hat{p}'_{(1,2)}\big)\hat{T}^{\dagger},\qquad \hat{T}=e^{i\hat{F}/\hbar}. 
\end{align}
As an example, we consider $\mathcal{O}=\mathcal{P}_1$, where $\mathcal{P}_1$ is the mechanical momentum of particle 1. In $L$, we have
\begin{align}\label{eq:MechMomL}
\mathcal{P}_1\xrightarrow[]{L}\hat{O}\big(\hat{x}_{(1,2)},\hat{p}_{(1,2)}\big)=\hat{p}_1\stackrel{\text{pos.}}{=}-i\hbar\partial_{x_1},
\end{align}
where `pos.' stands for the position space representation. Instead, following the prescription in Eq.~\eqref{eq:Unitary}, $\mathcal{P}_1$ in $L'$ is given by 
\begin{align}\label{eq:MechMomL'}
\mathcal{P}_1\xrightarrow{L'}\hat{T}\hat{p}'_1\hat{T}^{\dagger} = \hat p_1' - \partial \hat F/\partial{x_1}\stackrel{\text{pos.}}{=}-i\hbar\partial_{x_1}-\partial F/\partial{x_1}.
\end{align}
It is in this sense that the classical relation~\eqref{eq:ConjMom} must be understood quantum mechanically. Note that this relation is a bit more subtle to express after quantization, because even though they represent different physical quantities, $\hat{p}_1$ and $\hat{p}'_1$ have the same position space representation $-i\hbar\partial_{x_1}$ in their respective Lagrangian descriptions $L$ and $L'$ \cite{TannoudjiPartOneChapterFour}. We refer to Table~\ref{table:tarns} where we present how different observables of interest are expressed in $L$ and $L'$.

The same unitary transformation also relates the statevectors in the two descriptions. If $\ket{\psi}$ satisfies the relation
\begin{align}\label{eq:StateL}
\hat{O}\ket{\psi} = \psi \ket{\psi},
\end{align}
the corresponding eigenvector for $\hat{O}'$, with the same eigenvalue, is given by 
\begin{align}\label{eq:StateL'}
\hat{O}'\ket{\psi}' = \psi \ket{\psi}',\qquad \ket{\psi}'=\hat{T}\ket{\psi}.
\end{align}
Therefore, although the two quantum descriptions appear to be  different at first sight, the operator $\hat{T}$ allows one to map the calculations performed for one Lagrangian to those performed for the other.
More importantly, following the correspondence, the expectation values coincide: 
\begin{align}
\langle \psi(t) |\hat{O}  |\psi(t)\rangle_t ={} '\langle\psi(t) |\hat{O'}  |\psi(t) \rangle'_t.
\end{align}
The crucial point is that, while such a map exists globally (for S+E), and makes the two descriptions equivalent, it does not hold locally at the level of the reduced density matrix $\hat{\rho}_1$ of the system alone, whenever $\hat F$ involves variables of both the system (particle 1) and the environment (particle 2). This is the case in the previous example where $\hat{F} = -g\hat{x}_1\hat{x}_2$ (cf.~Eq.~\eqref{eq:Ftoy}), and also for QED, as we will see later. (One encounters the same situation in linearized gravity. See, for example, the interaction term in the action for linearized gravity \cite{Wilczek2021}.) 

The global nature of $\hat{F}$ leads to two different master equations~\eqref{eq:Master_L1} and~\eqref{eq:Master_L2}. The origin of this difference can be split into two parts, which we now discuss: the choice of the initial state and the difference between the conjugate momenta. 

\subsection{The initial state}
For the toy model described in Sec.~\ref{sec:InEq}, the master equations~\eqref{eq:Master_L1} and~\eqref{eq:Master_L2} are  obtained starting from the same factorized initial state, as customary (although not necessary) in open quantum systems. However, this choice of initial state for the two Lagrangians is incompatible with Eqs.~\eqref{eq:StateL} and~\eqref{eq:StateL'}. If a two-particle state is taken to be $\psi(x_1,x_2)$ in $L$, the state encoding the same physics is given by $e^{-igx_1x_2/\hbar}\psi(x_1,x_2)$ in $L'$. Consequently, if calculations are performed with a factorized initial System-Environment (S-E) state in one representation, i.e. 
\begin{align}
\rho(\tilde{x}_{(1,2)},x_{(1,2)})= \rho_1(\tilde{x}_1,x_1)\rho_2(\tilde{x}_2,x_2),
\end{align}
one should start with a non-factorized initial state 
\begin{align}
\rho= \rho_1(\tilde{x}_1,x_1)e^{\pm ig\tilde{x}_1\tilde{x}_2/\hbar}\rho_2(\tilde{x}_2,x_2)e^{\mp igx_1x_2/\hbar}
\end{align}
in the other representation to describe the same physics.

This implies that the customary factorized initial state assumption, inevitably leads to two different master equations, as can be seen from the fact that the expectation values $\langle\hat{x}_1\rangle_t$, when obtained from Eqs.~\eqref{eq:Master_L1} and~\eqref{eq:Master_L2}, do not match, as detailed in Appendix~\ref{appendix:B}. This discrepancy due to the choice of the same factorized initial state in both representations is not only important for the toy model we discuss, but also because the same choice is made in the literature for describing non-relativistic QED, independently of the representation of the interaction term, $\mathbf{A}\cdot\dot{\x}$ \cite{AnastopoulosQED,BandP} or $\mathbf{E}\cdot\x$ \cite{Caldeira1991, Gundhi:2023vjs}.

Nevertheless, we emphasize that while the same factorized initial state leads to different master equations,  it does not pose a fundamental problem. Even though, in general, it is much more challenging to perform calculations for a non-factorized initial state, it can be done in principle. We show in Appendix~\ref{appendix:C} how the expectation values for $\langle\hat{x}_1\rangle_t$ coincide  if, for instance, the calculations are performed with the corresponding non-factorized initial state in $L$. 

We now discuss the second source of discrepancy, namely, the difference between  $\hat{p}_1$ and $\hat{p}'_1$. Unlike the initial state, as we shall point out, the difference between $\hat{p}_1$ and $\hat{p}'_1$ irreversibly breaks the equivalence between the two master equations.

\subsection{The system variables $\hat{p}_1$ and $\hat{p}'_1$}

The fundamental difference between the master equations~~\eqref{eq:Master_L1} and~\eqref{eq:Master_L2} is that they are functions of different physical operators, which are not related locally. Specifically,  
the Lagrangian $L$ assigns the mechanical momentum $\hat{p}_1$ to particle 1, while $L'$ assigns $\hat{p}'_1$. As explained before, the two operators are different; as a matter of fact,  Eqs.~\eqref{aeq:EOML'1_int} and~\eqref{aeq:EOML_Trans}  show that the expectation value $\langle\hat{p}_1'\rangle_t$ derived from $L'$, and $\langle\hat{p}_1\rangle_t$ derived from $L$ with an appropriately transformed initial state, do not match. 

Let us consider the observable $\mathlarger{\mathlarger{\pi}}_1$ corresponding to the conjugate momentum of particle 1 in $L'$; upon quantization:
\begin{align}\label{eq:ConjMomL'}
\mathlarger{\mathlarger{\pi}}_1\xrightarrow[]{L'}\hat{p}'_1\stackrel{\text{pos.}}{=}-i\hbar\partial_{x_1}.
\end{align}
The central issue is that the same observable 
 is represented by a nonlocal operator in $L$:
\begin{align}\label{eq:ConjMomL}
\mathlarger{\mathlarger{\pi}}_1\xrightarrow{L}\hat{T}^{\dagger}\hat{p}_1\hat{T} = \hat p_1 + \partial \hat F/\partial{x_1}\stackrel{\text{pos.}}{=}-i\hbar\partial_{x_1}+\partial F/\partial{x_1}.
\end{align}
While $\hat{p}'_1$ is local to the system in $L'$, it is represented by the global S+E  operator $\hat{p}_1-g\hat{x}_2$ in $L$. The opposite is true for $\mathcal{P}_1$, as shown before in Eqs.~\eqref{eq:MechMomL} and~\eqref{eq:MechMomL'}.

The conclusion is that observables represented locally in one representation are represented globally in the other representation, since the unitary transformation $\hat{T} = e^{-ig\hat{x}_1\hat{x}_2/\hbar}$ involves both the system and environment operators. 

This poses the following difficulty. Let us consider, for example, whether there is decoherence in the eigen basis of (the operator representation of) $\mathlarger{\mathlarger{\pi}}_1$. The master equation~\eqref{eq:Master_L2}, derived from $L'$, says yes. However, this question cannot be sensibly answered by looking at only the master equation~\eqref{eq:Master_L1}, which is derived from $L$.

For the same reason, the master equations derived from $L$ and $L'$ cannot be related by the same unitary transformation $\hat{T} = e^{-ig\hat{x}_1\hat{x}_2/\hbar}$. This is because, by definition, they are obtained by tracing over the environment (particle 2). Global S+E operators such as $\hat{T}$ cannot be present at the level of the reduced master equation in either of the two Lagrangian descriptions, and thus cannot be used in going from Eq.~\eqref{eq:Master_L1} to Eq.~\eqref{eq:Master_L2} directly. 

The two reduced density matrices are separately valid as long as one is interested in either functions of the conjugate variables $(\hat x_1, \hat p_1)$ or $(\hat x_1, \hat p_1')$. But one cannot switch from one to the other---unless one goes back to the full density matrix---because the partial trace and the transformation $\hat T$ are incompatible with each other: once the partial trace has been taken, the transformation $\hat T$ cannot be applied anymore.

The question then becomes to decide which set of conjugate variables, and therefore which representation, to use. In theory, any pair does the job. But from the {\it operational} point of view there is a difference. 
If the objective is to describe the reduced dynamics of particle 1, one would typically not be interested in the observable $\mathlarger{\mathlarger{\pi}}_1$. This is because, operationally, measuring $\mathlarger{\mathlarger{\pi}}_1$ not only requires measuring the velocity of particle 1, but also the position of particle 2.

While the conjugate momentum $\hat p_1$ coincides with the mechanical momentum $m_1 \dot {\hat x}_1$, and therefore it can be determined by knowing the mass and by measuring the velocity of particle 1, the conjugate momentum $\hat p_1'$ corresponds to the mechanical momentum minus $g \hat x_2$ and its measurement passes through monitoring also the position of particle 2, which however is supposed not to be accessible at the level of the system (particle 1) alone. Then, having decided at the beginning that only particle 1 is accessible and particle 2 is traced away, the only consistent choice from the operational point of view is to use the representation in terms of the conjugate variables $(\hat x_1, \hat p_1)$. This choice returns a master equation that is similar to the Caldeira-Leggett master Eq.~\eqref{eq:cl}; this would not be the case with the other choice \cite{AnastopoulosQED,BandP_Brem2001,BandP}. 

This does not imply that, in general, the $L$ representation is better than $L'$, for describing open quantum systems. If, for instance, we were interested in the effective dynamics of particle 2, after averaging over particle 1, following the same line of reasoning, $L'$ would be operationally better than $L$ to derive the master equation for particle 2.

At a general level, the main takeaway message of this analysis is that a global unitary transformation changes the S-E separation for any bi- or multi-partite system (simplest example being that of two spin-half particles), and would thus lead to inequivalent reduced dynamics. The inference on which variables, before or after the unitary transformation, appropriately describe the open dynamics of the system of interest must be done from physical considerations, and keeping in mind that certain operators which are local to the system, (e.g. $\hat{p}'_1$) do not necessarily represent system observables that can be measured independently of the environment.

In the next section, we apply these considerations to a particle interacting with the electromagnetic field. As can be immediately seen by comparing the QED Lagrangian with the toy model discussed before, $x_1$ is analogous to the position of the charged particle in the QED Lagrangian, while $x_2$  is analogous to the vector potential of the electromagnetic field.

\begin{table}
\centering
\begin{tabularx}{0.48\textwidth} { 
  |>{\centering\arraybackslash}X  
  | >{\centering\arraybackslash}X 
  | }
 \hline
 {\bf Lagrangian $L$} & {\bf Lagrangian $L'$} \\
 \hline   
 \multicolumn{2}{|c|}{\bf Full systems S+E} \\
 \hline 
 \multicolumn{2}{|c|}{\bf Operators} \\
\hline
Conj. var.: $(\hat x_1, \hat p_1)$, $(\hat x_2, \hat p_2)$ & Conj. var.: $(\hat x_1, \hat p_1')$, $(\hat x_2, \hat p_2')$ \\
\hline 
\multicolumn{2}{|c|}{Position of particle 1 (position representation)} \\ \hline
 $\hat{x}_1$ ($x_1$)& $\hat{x}_1$ ($x_1$)\\
 \hline
\multicolumn{2}{|c|}{Position of particle 2} \\ \hline
 $\hat{x}_2$ ($x_2$) & $\hat{x}_2$ ($x_2$)\\
 \hline
\multicolumn{2}{|c|}{Mechanical momentum of particle 1} \\ \hline
  $\hat{p}_1$ ($-i\hbar\partial_{x_1}$) & $\hat{p}'_1+g\hat{x}_2$  ($-i\hbar\partial_{x_1} +gx_2$)\\
  \hline
\multicolumn{2}{|c|}{Mechanical momentum of particle 2} \\ \hline
$\hat{p}_2 - g\hat{x}_1$  ($-i\hbar\partial_{x_2} -gx_1$) & $\hat{p}'_2$ ($-i\hbar\partial_{x_2}$)\\
 \hline
\multicolumn{2}{|c|}{\bf States} \\
\hline
$\ket{\psi}$ & $\ket{\psi}'=\hat T \ket{\psi}$ \\
 pos.: $\psi(x_1, x_2) $ &  pos.: $ e^{-igx_1x_2/\hbar}\,\psi(x_1,x_2)$\\
 \hline  
\multicolumn{2}{|c|}{\bf Reduced system S (trace over E)} \\
 \hline
 $\hat \rho_1 = \text{Tr}^{(2)} [\hat \rho] = \hat \rho_1(\hat x_1, \hat p_1)$ & $\hat \rho_1' = \text{Tr}^{(2)} [\hat T \hat \rho \hat T^\dagger] \!=\! \hat \rho_1'(\hat x_1, \hat p_1')$ \\
 \hline
\multicolumn{2}{|c|}{\it $\hat \rho_1$ and $\hat \rho_1'$ cannot be transformed into one another} \\
 \hline 
 \multicolumn{2}{|c|}{\bf Which representation to choose, operationally?} \\
 \hline
$x_1$ and $\hat p_1 = m_1 \hat v_1$ can be measured only through S & $\hat p_1' = m_1 \hat v_1 - g \hat x_2$ requires controlling both S and E\\
 \hline
\end{tabularx}
\caption{Various quantum mechanical entities and their representation in terms of the conjugate variables of $L$ and $L'$. At the global S+E level, the operators and the statevectors in $L$ and $L'$ are related to each other by the unitary transformation $\hat{T}=\exp{-ig\hat{x}_1\hat{x}_2/\hbar}$. However, since $\hat{T}$ involves both the system ($\hat{x}_1$) and environment ($\hat{x}_2$) degrees of freedom, some observables which are local to the system when represented in $L$, correspond to non-local entities when expressed in $L'$: while position of particle 1 remains a local to the system in both $L$ and $L'$, the mechanical momentum of particle 1, which is local to the system in $L$ (it acts trivially on the environment basis states), corresponds to a global S+E operator for Lagrangian $L'$ (acting non-trivially on both the system and environment basis states). The opposite is true for the mechanical momentum of particle 2. Similarly, a factorized state in $L$, in general, corresponds to a non-factorized state in $L'$ (and vice versa). Since locality to the system is not preserved in going from $L$ to $L'$, and due to the non-unitary trace operation, $\hat{\rho}_1$ and $\hat{\rho}'_1$ are not related by $\hat{T}$. The unprimed representation must be preferred operationally for deriving the reduced dynamics of particle 1, since therein, all physically relevant system variables are accessible at the level of $\hat{\rho}_1$. This is not the case for $\hat{\rho}'_1$, since in $L'$, the velocity operator of particle 1 is expressed in terms of $\hat{x}_2$, which is traced out and not accessible at the level of $\hat{\rho}'_1$.}\label{table:tarns}
\end{table}

\section{Bremsstrahlung and open system dynamics for a non-relativistic charged particle}
In the Coulomb gauge, the Lagrangian describing the dynamics of a charged particle traveling at non-relativistic speeds ($v\ll c$) and interacting with the electromagnetic (EM) field is given by (cf.~\cite{TannoudjiPartOneChapterTwo} and Appendix A in \cite{Gundhi:2023vjs})
\begin{align}\label{eq:LQED}
L = \frac{1}{2}m\dot{\r}_e^2-V_0(\r_e)+\int\!d^3r \mathcal{L}_{\text{\tiny{EM}}}+e\r_{e}\dot{\mathbf{A}}_{\perp}\,.
\end{align}
Here, $\r_e$ denotes the position of the charged particle, $V_0$ is some external potential that acts only on the charged particle (system of interest), $\mathcal{L}_{\mathrm{EM}}$ is the Lagrangian density for the free EM field, and $\mathbf{A}_{\perp}$ is the transverse vector potential.

Alternatively, one can consider the Lagrangian $L'$ equal to $L$, with only the interaction term replaced by $-e \dot{\r}_e\mathbf{A}_{\perp}$; these two choices are well known in the literature as the $\bf{E} \cdot {\bf x}$ coupling and $\bf{A} \cdot {\bf p}$ coupling respectively \cite{TannoudjiPartOneChapterFour}. 
The two Lagrangians are of course equivalent, because they differ by a total derivative: $L'=L+dF_{\text{\tiny{EM}}}/dt $, where $F_{\text{\tiny{EM}}} = -e\r_e\mathbf{A}_{\perp}$; note that, as in the toy model described in Sec.~\ref{sec:InEq}, $F_{\text{\tiny{EM}}}$ involves both the system and the environmental variables. 

Our goal is to derive the master equation for the electron by tracing out the electromagnetic degrees of freedom. In the literature, to analyze decoherence due to vacuum fluctuations, both approaches---starting from $L$ \cite{Caldeira1991,Gundhi:2023vjs} and from $L'$ \cite{AnastopoulosQED,BandP}---have been considered, leading to contrasting results. As discussed earlier, this discrepancy is to be expected. Indeed, as previously noted, there is only one operationally viable choice: the one in which the conjugate momentum to ${\r}_e$ coincides with the mechanical momentum $m \dot {\r}_e$. This amounts to choosing the $\bf{E} \cdot {\bf x}$ coupling given by $L$ \footnote{For $L'$, the conjugate momentum to ${\r}_e$ is $m\dot{\r}_e-e\mathbf{A}_{\perp}(\r_e)$.}; we proceed with this choice to analyze decoherence due to bremsstrahlung. The Hamiltonian corresponding to the Lagrangian $L$ is 
given by
\begin{align}\label{eq:HQED}
H= \frac{\p^2}{2m}+{V}_0+V_{\text{\tiny{EM}}}+ \int\!d^3r \mathcal{H}_{\text{\tiny{EM}}}+ e\r _e\mathbf{\Pi}_{\text{\tiny{E}}}\,.
\end{align}
Here, $\mathcal{H}_{\text{\tiny{EM}}}$ represents the Hamiltonian density corresponding to the free EM field, $V_{\text{\tiny{EM}}}(\r_e):= e^2k^3_{\text{\tiny{max}}}\r^2_e/(3\pi^2\epsilon_0)$ and $V_0$, as mentioned before, is an external potential which we take to be the harmonic potential $V_0 =m\Omega^2 \hat{x}^2/2$. The derivation of the Hamiltonian in Eq.~\eqref{eq:HQED} starting from Eq.~\eqref{eq:LQED} can be found in Appendix A in \cite{Gundhi:2023vjs}. The reduced dynamics can be obtained by applying the influence functional formalism \cite{Feynman_Vernon}. 

It is well-known that an accelerated electron emits photons. These photons, emitted at different points along the electron trajectory, carry which-path information and are thus expected to lead to decoherence in position. Furthermore, the energy carried away by the photons would lead to a dissipative term in the equation of motion for the electron.

As for the toy model discussed in Sec.~\ref{sec:InEq}, the master equation for the electron interacting with the EM field can be obtained via the influence functional formalism, as introduced in Appendix~\ref{appendix:B}. Description of the formalism can also be found in \cite{calzetta_hu_2008} and \cite{Gundhi:2023vjs} -- where it was applied specifically to QED. To leading order, the master equation is obtained to be:
\begin{align}\label{eq:MasterEquation}
\partial_t\hat{\rho}_r=&\!-\!\frac{i}{\hbar}\!\left[\hat{H}_s,\hat{\rho}_{r}(t)\right]\!-\!\frac{1}{\hbar}\!\int_{0}^{t}\! d\tau \mathcal{N}(\tau)\left[\hat{x},\left[\hat{x}_{\text{\tiny{H}}_s}(-\tau),\hat{\rho}_{r}(t)\right]\right]\nonumber\\
&+\frac{i}{2\hbar}\int_{0}^{t}\!
d\tau\mathcal{D}(\tau)\left[\hat{x},\{\hat{x}_{\text{\tiny{H}}_s}(-\tau),\hat{\rho}_{r}(t)\}\right]\,.
\end{align} 
Here
\begin{align}
\hat{H}_s:=\hat{p}^2/(2m)+\hat{V}_0+\hat{V}_{\text{\tiny{EM}}}\,,
\end{align}
and 
\begin{align}
\hat{x}_{\text{\tiny{H}}_s}(-\tau):= \hat{U}^{-1}_s(t-\tau;t)\hat{x}\hat{U}_s(t-\tau;t)\,,
\end{align} 
with $\hat{U}_s(t-\tau;t)$ being the unitary operator that evolves the statevector of the system from time $t$ to $t-\tau$ via the system Hamiltonian $\hat{\mathrm{H}}_s$ and  $\hat{x}$ is the usual Schr\"{o}dinger operator such that $\hat{x}_{\text{\tiny{H}}_s}(0) = \hat{x}$. We adopt the convention where the electron moves along the x axis. 

Computation of the noise kernel requires the specification of the initial state. We assume a factorized system-environment initial state, and the EM field to be initially in its vacuum state $\hat{\rho}_{\mathrm{EM}}(t_i) = \ket{0}\bra{0}$. For such an initial state and the $\hat{x}\hat{{\Pi}}^{x}_{\mathrm{E}}$ interaction, the noise and dissipation kernel were computed explicitly in \cite{Gundhi:2023vjs}. They read:
\begin{align}\label{eq:NoiseVac}
\mathcal{N}_{0}(\tau)= \frac{4\alpha\hbar}{\pi c^2}\frac{\left(\epsilon^4-6\epsilon^2\tau^2+\tau^4\right)}{\left(\epsilon^2+\tau^2\right)^4}\,,
\end{align}
and:
\begin{align}\label{eq:DissipationKernelOne}
\mathcal{D}(\tau)&=\frac{4\hbar\alpha}{3c^2}\theta(\tau)\delta'''_{\epsilon}(\tau)\,,
\end{align} 
where $\alpha = e^2/(4\pi\epsilon_0\hbar c)$ is the fine-structure constant, $\delta_{\epsilon}:=\frac{1}{\pi}\frac{d}{d\tau}\tan^{-1}\left(\frac{\tau}{\epsilon}\right)$, and $\epsilon = 1/\Omega_\text{max}$ is a high energy cutoff to regularize the otherwise divergent dynamics. 

The master equation~\eqref{aeq:MasterEq}, with the noise and dissipation kernels in Eqs.~\eqref{eq:NoiseVac} and~\eqref{eq:DissipationKernelOne} respectively, is the one that would be obtained to second order for a particle interacting with the EM field initially in the ground state. Nevertheless, the integrals appearing in the master equation depend on the free evolution of the position operator and are thus specific to the physical scenario under consideration.  We now proceed towards the derivation of the master equation, specific to bremsstrahlung.

To bring the master equation closer to the Caldeira-Leggett form, the integrals involving the noise and the dissipation kernels must be computed. We first focus on the dissipation kernel. 
From Eq.~\eqref{eq:DissipationKernelOne}, it can be seen that it is proportional to the third derivative of the Dirac delta $\delta_{\epsilon}$. The integral involving the dissipation kernel can be computed by integrating by parts. Using the standard properties of the Dirac delta, i.e., $\delta'_{\epsilon}(0)=0$, $\delta'''_{\epsilon}(0)=0$ (since $\delta'_{\epsilon}(t)$ and $\delta'''_{\epsilon}(t)$ are odd functions), and $\delta_{\epsilon}(t)=0$ for $t\gg\epsilon$, one arrives at the relation $\int_{0}^t\!d\tau \delta'''_{\epsilon}(\tau)f(-\tau) = \left.\left(f'''/2 - \delta_{\epsilon}f''-\delta''_{\epsilon}f\right)\right|_{\tau=0}$.  Substituting the Heisenberg operator $\hat{x}_{\text{\tiny{H}}_s}(-\tau)$ for the general function $f$, and using the standard relations for the harmonic oscillator: $\hat{x}_{\text{\tiny{H}}_s}(0)=\hat{x}$, $\hat{x}''_{\text{\tiny{H}}_s}(0)=-\Omega^2\hat{x}$, $\hat{x}'''_{\text{\tiny{H}}_s}(0)=-\Omega^2\hat{p}/m$, the master equation~\eqref{eq:MasterEquation} becomes:
\begin{align}\label{AEQ:MEQ}
\dot{\hat{\rho}}_r = &-\frac{i}{\hbar}\left[\frac{\hat{p}^2}{2m}+\frac{1}{2}m\Omega^2\hat{x}^2,\hat{\rho}_r\right]-\frac{i\alpha\Omega^2}{3mc^2}\left[\hat{x},\{\hat{p},\hat{\rho}_r\}\right]  \nonumber\\
&-\frac{1}{\hbar}\int_{0}^{t}\! d\tau \mathcal{N}_0(\tau)\left[\hat{x},\left[\hat{x}_{\text{\tiny{H}}_s}(-\tau),\hat{\rho}_{r}\right]\right]\,.
\end{align}
The contribution due to $\hat{V}_\text{\tiny{EM}}$, which appeared in the Hamiltonian~\eqref{eq:HQED}, is canceled exactly by $\delta''_{\epsilon}(0)$, while the $\delta_{\epsilon}(0)$ term that appears after integration by parts leads to the standard renormalization of mass/frequency, which we have included in the redefinition of the parameters  (cf.~Appendix~\ref{appendix:D}) \footnote{It is worth emphasizing that the Lagrangian~\eqref{eq:LQED} naturally provides the so-called counterterm $V_\text{\tiny{EM}}$ which would have to be added by hand in the phenomenological models with a supraohmic bath \cite{Caldeira1991}.}.

We now study {\it decoherence} and compute the integral involving $\mathcal{N}_{0}$. The contribution of $\mathcal{N}_0$ to the master equation splits into two pieces:
\begin{align}
\left.\partial_t\hat{\rho}_{r}\right|_{\mathcal{N}_0}= &-\frac{1}{\hbar}\int_{0}^{t} d\tau \mathcal{N}_{0}(\tau)\cos{\Omega \tau}\left[\hat{x},\left[\hat{x},\hat{\rho}_{r}(t)\right]\right]\ \nonumber\\
&+\frac{1}{ m\hbar\Omega}\int_{0}^{t} d\tau \mathcal{N}_{0}(\tau)\sin{\Omega \tau}\left[\hat{x},\left[\hat{p},\hat{\rho}_{r}(t)\right]\right]\,.
\end{align}
Both terms have a complicated time dependence, but they can be evaluated in the physically relevant regime $t\gg\epsilon$. In particular, the first line can be computed by noting that:
\begin{align}\label{AEQ:N0reg}
\frac{1}{\hbar}\int_{0}^{t} d\tau \mathcal{N}_{0}(\tau)\cos{\Omega \tau} &= \frac{1}{2\hbar}\mathfrak{Re} \int_{-t}^{t}d\tau \mathcal{N}_{0}(\tau)e^{i\Omega\tau} \nonumber\\
\overset{t\rightarrow\infty}&{=}\sqrt{\frac{\pi}{2\hbar^2}}\mathfrak{Re} \{\mathfrak{F}[\mathcal{N}_0] (\Omega)\} \nonumber\\ &=\frac{\alpha \Omega^3}{3c^2}e^{-\epsilon\Omega}\,,
\end{align}
where $\mathfrak{F}$ denotes the Fourier transform. The term in Eq.~\eqref{AEQ:N0reg} is well behaved for an arbitrarily small $\epsilon$ (or an arbitrarily large cutoff $\Omega_{\text{max}}$, $\epsilon=1/\Omega_{\text{max}}$, such that $\Omega/\Omega_{\text{max}}\ll 1$).  At late times, in the limit $\epsilon\rightarrow0$, this term contributes to the master equation as:
\begin{align}
\left.\partial_t\hat{\rho}_r\right|_{\text{dec}}= -\frac{\alpha\Omega^3}{3c^2}\left[\hat{x},\left[\hat{x},\hat{\rho}_{r}(t)\right]\right]\,,
\end{align}
and describes decoherence in position. The second term, in the limit $t\gg \epsilon$, can be expanded to:
\begin{align}\label{eq:N0Final}
&\frac{1}{m\hbar\Omega}\int_{0}^{t} d\tau \mathcal{N}_{0}(\tau)\sin{\Omega\tau}\overset{t\to\infty}{=}\nonumber\\
&\left(-\frac{2\alpha}{3\pi m c^2\epsilon^2}  + \frac{2 \alpha\Omega^2}{
	3\pi mc^2}\log(\epsilon \Omega)+ \frac{2\alpha\gamma_\text{\tiny{E}}\Omega^2}{
	3\pi m c^2}\right) + O(\epsilon)\,,
\end{align}
where $\gamma_\text{\tiny{E}}$ is the Euler-Mascheroni constant. The first term is independent of $\Omega$ and scales explicitly with the UV cutoff $1/\epsilon$. 
It has been argued and shown in \cite{Diosi1995,Unruh_Coherence,Gundhi:2023vjs,Gundhi2024} that in tracing over the degrees of freedom of the EM field, one also, unphysically, traces over the dressed states of the charged particle. The way that the vacuum of the EM field reacts to the presence of a free charged particle is by dressing the bare particle with a cloud of virtual photons \cite{Greiner1996}. A charged particle is typically observed with its dressing, and the dressed states must therefore be considered part of the system and not the environment \cite{Diosi1995}. Tracing over them describes false decoherence that would not be observed in a typical interference experiment. The first term in Eq.~\eqref{eq:N0Final} survives even for a free particle when $\Omega=0$, and corresponds to an effect due to tracing over the particle dressing. We take it to be unphysical and therefore discard it.

The second term in Eq.~\eqref{eq:N0Final}, which diverges logarithmically, has been  discussed before, for example, in \cite{Agon2017}.
There, it was argued that such a divergence leads to the renormalization of the two point function $\langle p(t_1)p(t_2)\rangle$  in the limit $t_1\rightarrow t_2$.
However, since our calculations are performed in the non relativistic regime, we take the point of view  that the value of the cutoff must be bounded from above such that $\hbar \Omega_\text{max} \lesssim mc^2$ (c.f. pgs. 200-202 in \cite{TannoudjiPartOneChapterThree}), which makes the logarithmic piece finite in any case. Furthermore, the contributions coming from this regularized logarithmic piece as well as the last term in Eq.~\eqref{eq:N0Final},  which is independent of the cutoff, are suppressed in the non relativistic limit $\hbar\Omega \ll mc^2$. This can be seen by looking at the asymptotic variances presented in Appendix~\ref{appendix:D}.

Therefore, retaining only the physically relevant contributions, the master equation for the harmonically trapped electron emitting bremsstrahlung reads:
\begin{align}\label{eq:meqBrem}
\partial_t\hat{\rho}(t) =&-\frac{i}{\hbar}\left[\frac{\hat{p}^2}{2m}+\frac{1}{2}m\Omega^2\hat{x}^2,\hat{\rho}_{r}\right]\nonumber\\
&-\frac{i\alpha\Omega^2}{3mc^2}\left[\hat{x},\{\hat{p},\hat{\rho}_r\}\right] -\frac{\alpha\Omega^3}{3c^2}\left[\hat{x},\left[\hat{x},\hat{\rho}_{r}(t)\right]\right]\,.
\end{align}
The first term on the RHS of the master equation~\eqref{eq:meqBrem} is the Liouville-von Neumann term, which describes how the density matrix for a harmonic oscillator (with a renormalized frequency) would evolve, in the absence of any environment. The second term describes a friction term in the equation of motion for $d\langle\hat{p}\rangle_t/dt$, and thus describes the energy lost by the oscillating charged particle due to photon emission. Finally, the last term describes decoherence in position, which can be intuitively understood as being due to which-path information carried away by radiation emitted by the charged particle.

Thus, we have shown that if the canonical variables assigned to the system are the ones obtained from $L$, rather than $L'$, one gets decoherence in the position basis and that the dynamics of the reduced density matrix closely resembles the well-known Caldeira-Leggett master equation.

\section{Conclusions} In this work we have shown that two Lagrangians $L$ and $L'$, that differ by a total derivative and are therefore unitarily equivalent to each other globally, give different predictions at the level of reduced system dynamics, due to the different canonical variables they ascribe to the system and environment. 

With the help of a toy model, we show that the master equation for the reduced density matrix of the system may predict  decoherence in position or (canonical) momentum, depending upon whether the dynamics is derived from $L$ or $L'$ respectively, and that the two master equations in general are not related to each other. We then show that the same situation would be encountered for the physical QED Lagrangian describing the dynamics of the non-relativistic charged particle.

While there are several works which derive open system dynamics starting from a given Lagrangian \cite{UnruhZurek1989,HuPazZang1992,BandP,AnastopoulosQED,BandP_Brem2001}, in this work we discuss why a certain choice of Lagrangian would be more suitable for describing the loss of which-path information, as it ascribes conjugate variables to the system which are operationally more meaningful. Following this reasoning, we break the degeneracy between the otherwise equivalent QED Lagrangians where the interaction terms are given by ${\bf r}\cdot\dot{\bf{A}}_{\perp}$ or $\dot{{\bf r}}\cdot\bf{A}_{\perp}$. 

Then, performing a microscopic derivation of the reduced dynamics for an accelerated charged particle, starting from the $\mathbf{r}\cdot\dot{\mathbf{A}}_{\perp}$ Lagrangian, we show that the master equation closely resembles  the phenomenological Caldeira-Leggett equation and predicts decoherence in the position basis. This contrasts the result obtained in \cite{BandP_Brem2001}, where the master equation for bremsstrahlung predicts decoherence in the conjugate momentum basis, and was derived starting from the $\dot{\mathbf{r}}\cdot\mathbf{A}_{\perp}$ Lagrangian.

Given the ambiguity, we put forward the point of view  that while both the master equations might be mathematically consistent, decoherence in position basis is more reasonable to expect from phenomenological considerations, and in fact follows from a microscopic derivation of the reduced dynamics, provided one picks the QED Lagrangian that assigns operationally meaningful conjugate variables to the system. The present analysis might also be relevant to other systems where an effective Hamiltonian is used to describe, for instance,  the interaction between two-level systems or that between a two-level system and a cavity mode. As it is the case for a point charge, a unitary transformation which mixes the S-E separation for discrete systems would also lead to inequivalent effective system dynamics in general. Given the inequivalence, the correct system variables must be chosen for describing the effective system dynamics, without assuming a-priori that all conjugate variables which mathematically belong to the system-Hilbert space are meaningful system variables. Such a choice, depending on the scenario being studied, might be possible to make at the level of the effective Hamiltonian itself, or by resorting to the QED Hamiltonian from which the toy/effective Hamiltonians are derived, for describing various scenarios involving matter-light interaction.

\section*{Acknowledgements}
We acknowledge financial support from the University of Trieste, INFN and the EIC Pathfinder project QuCoM (GA No. 101046973). A.B. also acknowledges financial support from the PNRR MUR project PE0000023-NQSTI.

\appendix

\section{Influence functional formalism and the master equation for the toy model}\label{appendix:B} Within the path integral formalism, the density matrix at time $t$ can be written as \cite{calzetta_hu_2008}
\begin{align}\label{aeq:DensityMatrixGeneral}
\bra{\tilde{x}^{\text{\tiny{f}}}}\hat{\rho}(t)\ket{x^{\text{\tiny{f}}}}  ={}& \int_{\substack{{x(t) = x^{\text{\tiny{f}}} ,}\\ {\tilde{x}(t)= \tilde{x}^{\text{\tiny{f}}}}}}D[x,p,\tilde{x},\tilde{p}]e^{\frac{i}{\hbar}(S[\tilde{x},\tilde{p}]-S[x,p])}\rho^i\,,
\end{align}
where $\rho^i:=\rho(\tilde{x}_i,x_i,t_i)$, $S[x,p] := \int_{t_i}^{t}dt (p\dot{x}-H[x,p])$, and the integrals over $x_i$ and $\tilde{x}_i$ are included within the path integral.

In the toy model described in Sec.~\ref{sec:InEq}, there are two particles. The full density matrix is therefore obtained by doubling the degrees of freedom in Eq.~\eqref{aeq:DensityMatrixGeneral} as $x\rightarrow x_{(1,2)}$ and $p\rightarrow p_{(1,2)}$ (and likewise for $\tilde{x}$ and $\tilde{p}$).

From the full density matrix, the reduced density matrix is typically obtained by following the standard prescription:
\begin{enumerate}
    \item Assume a factorized initial condition such that $\rho^i(\tilde{x}_{(1,2)},x_{(1,2)})=\rho_{1}^i(\tilde{x}^i_{1},x^i_{1})\rho_{2}^i(\tilde{x}^i_{2},x^i_{2})$\, where  $\rho_{2}^i(\tilde{x}^i_{2},x^i_{2})$ is a Gaussian, mean zero state.
    \item Trace over the mathematical degrees of freedom assigned to particle 2.
\end{enumerate}

Before performing the trace, $S$ appearing inside the phase-space weighing function $\exp{iS/\hbar}$, which governs the full dynamics in Eq.~\eqref{aeq:DensityMatrixGeneral}, is decomposed as $S[\mu^{(1)},\mu^{(2)}]=$
\begin{align}
S_{1}[\mu_{1}]+S_{2}[\mu_{2}]+S_{\mathrm{int}}[\mu_{1},\mu_{2}].
\end{align}
Here, $\mu_{1}$ stands for the conjugate variables $\{x_{1},p_{1}\}$ assigned to particle 1 and similarly $\mu_{2}$ for conjugate variables of particle 2. 
For the dynamics described by $L$, from Eqs.~\eqref{eq:H_L} and~\eqref{aeq:DensityMatrixGeneral}, we get
\begin{align}
S_1^L &= \int_{t_i}^t dt \left(p_1\dot{x}_1-\frac{p^2_1}{2m_1}-\frac{g^2x^2_1}{2m_2}\right)\,,\nonumber\\
S_2^L&=\int_{t_i}^t dt \left(p_2\dot{x}_2-\frac{p^2_2}{2m_2}\right)\nonumber\,,\\
S^L_{\mathrm{int}}&=\int_{t_i}^t dt \frac{gx_1p_2}{m_2}\,.
\end{align}
Instead, when the dynamics is described by $L'$, from Eqs.~\eqref{eq:MasterEquation2} and~\eqref{aeq:DensityMatrixGeneral} we get
\begin{align}
S_1^{L'} &= \int_{t_i}^t dt \left(p'_1\dot{x}_1-\frac{p'^{2}_1}{2m_1}\right)\,,\nonumber\\
S_2^{L'} &= \int_{t_i}^t dt \left(p'_2\dot{x}_2-\frac{p'^{2}_2}{2m_2}-\frac{g^2x^2_2}{2m_1}\right)\,,\nonumber\\
S^{L'}_{\mathrm{int}}&=-\int_{t_i}^t dt \frac{gp'_1x_2}{m_1}\,.
\end{align}
The reduced density matrix $\hat{\rho}_r$ for particle 1 is obtained by tracing over the degrees of freedom for particle 2 such that
\begin{align}\label{eq:reduced_density}
\bra{\tilde{x}_1^{\text{\tiny{f}}}}\hat{\rho}_r(t)\ket{x_1^{\text{\tiny{f}}}}=&\int_{\substack{{x_{1}(t) = x_{1}^{\text{\tiny{f}}} ,}\\ {\tilde{x}_{1}(t)= \tilde{x}_{1}^{\text{\tiny{f}}}}}}D[\mu_{1},\tilde{\mu}_{1}]e^{\frac{i}{\hbar}(\tilde{S}_{1}-S_{1})}\rho^i_{1}\times\nonumber\\
&\int_{\mathrm{tr}}D[\mu^{(2)},\tilde{\mu}^{(2)}]e^{\frac{i}{\hbar}(\tilde{S}_{2}+\tilde{S}_{\mathrm{int}}-S_{2}-S_{\mathrm{int}})}\rho^{i}_2\,,
\end{align}
where 
\begin{align}
\int_{\mathrm{tr}}D[\mu^{(2)},\tilde{\mu}^{(2)}]:=\int dx_2(t)\int_{\substack{{x_2(t) = \tilde{x}_2(t)}}}D[\mu^{(2)},\tilde{\mu}^{(2)}]\,,
\end{align}
and $\tilde{S}$ and $S$ stand for $S[\tilde{\mu}]$ and $S[\mu]$ respectively. After tracing over the environment, the integral involving the environmental degrees of freedom $\mu_2$ in Eq.~\eqref{eq:reduced_density}  yields a Gaussian in the system degree of freedom $s$ \cite{calzetta_hu_2008}, which depends on the Lagrangian used to describe the dynamics. In particular, the trace over the environment yields a Gaussian in $s=x_1$ for $S_{\mathrm{int}}^{L}$ but a Gaussian in $s=p'_1$ for $S_{\mathrm{int}}^{L'}$. Therefore, 
\begin{align}
&\int_{\mathrm{tr}}D[\mu^{(2)},\tilde{\mu}^{(2)}]e^{\frac{i}{\hbar}(\tilde{S}_{2}+\tilde{S}_{\mathrm{int}}-S_{2}-S_{\mathrm{int}})}\rho_{2}^{i} =\nonumber\\ &\quad\exp{\frac{i}{2\hbar}\iint dt_1dt_2 M_{ab}(t_1;t_2)s^a(t_1)s^b(t_2)}\,.\label{Mab}
\end{align}
Here, we use the vector notation $s^a=s$ for $a=1$, $s^a=\tilde{s}$ for $a=2$ and $s_a=\eta_{ab}s^{b}$ with $\eta_{ab}=\mathrm{diag}(-1,1)$.

The matrix $M_{ab}$ determines the effective system dynamics, and how the the environment influences the system. The matrix elements can be obtained by acting with $\frac{\hbar}{i}\frac{\delta}{\delta s^{a}}\frac{\delta}{\delta s^{b}}|_{s^{a}=s^{b}=0}$ (where $s^a$ and $s^b$ are set to zero after taking the derivatives) on Eq.~\eqref{Mab} such that
\begin{align}\label{aeq:matrix_elements}
M_{ab}(t_1;t_2)=\frac{ig^2}{\hbar}\int_{\mathrm{tr}}E_{a}\left(t_1\right) E_{b}\left(t_2\right) e^{\frac{i}{\hbar}(\tilde{S}_{2}-S_{2})}\rho_{2}^{i}\,.
\end{align} 
As was the case with $s^a$, the degree of freedom $E^a$ depends on the Lagrangian used to describe the dynamics: $E^a = p^a_2/m_2$ for $L$ and $E^a = x^a_2/m_1$ for $L'$. 

Depending on  the indices $a$ and $b$, the matrix elements $M_{ab}(t_1;t_2)$ can be written as expectation values of the time-ordered ($\mathcal{T}$), anti-time ordered ($\tilde{\mathcal{T}}$), path-ordered or anti-path ordered products in the Heisenberg picture \cite{calzetta_hu_2008}. They are given by 
\begin{align}\label{eq:Mab}
&M_{ab}(t_1;t_2) =\nonumber\\ 
&\quad\frac{ig^2}{\hbar}\begin{bmatrix}
\left\langle\tilde{\mathcal{T}}\{ \hat{E}(t_1) \hat{E}(t_2)\}\right\rangle_0 & -\left\langle \hat{E}(t_1) \hat{E}(t_2)\right\rangle_0\\-\left\langle \hat{E}(t_2) \hat{E}(t_1)\right\rangle_0&\left\langle \mathcal{T}\{ \hat{E}(t_1) \hat{E}(t_2)\}\right\rangle_0 
\end{bmatrix}\,.
\end{align}
In Eq.~\eqref{eq:Mab}, the zero in the subscript denotes that the correlations of the environmental operator are calculated by disregarding the system-environment interaction.

In terms of the so-called influence functional $S_{\text{\tiny{IF}}}$ \cite{Feynman_Vernon},  the reduced density matrix in Eq.~\eqref{eq:reduced_density} can be written as 
\begin{align}\label{eq:RedRho_formal}
\bra{\tilde{x}_1^{\text{\tiny{f}}}}\hat{\rho}_{r}(t)\ket{x_1^{\text{\tiny{f}}}}= \int_{\substack{{x_1(t) = x_1^{\text{\tiny{f}}} ,}\\ {\tilde{x}_1(t)= \tilde{x}_1^{\text{\tiny{f}}}}}}D[\mu_1,\tilde{\mu}_1]e^{\frac{i}{\hbar}(\tilde{S}_{1}-S_1+S_{\text{\tiny{IF}}}[s,\tilde{s}])}\rho^i_1\,,
\end{align}
where 
\begin{align}
& S_{\text{\tiny{IF}}}[x,x^{\prime}] =\nonumber\\ 
&\quad\frac{1}{2}\int_{t_i}^{t} dt_1dt_2\begin{bmatrix}
s(t_1) &\tilde{s}(t_1)
\end{bmatrix}\cdot\begin{bmatrix}
M_{11}&M_{12}\\
M_{21}&M_{22}
\end{bmatrix}\cdot\begin{bmatrix}
s(t_2)\\\tilde{s}(t_2)
\end{bmatrix}\,.
\end{align}
Instead, in a different choice of basis $(X\,,u)$ defined by
\begin{align}
X(t):=&(\tilde{s}(t)+s(t))/2\,,\quad u(t)=\tilde{s}(t)-s(t)\,,
\end{align}
the influence functional reads 
\begin{align}
&S_{\text{\tiny{IF}}}[X,u] =\nonumber\\ 
&\quad\frac{1}{2}\int_{t_i}^{t} dt_1dt_2\begin{bmatrix}
X(t_1) &u(t_1)
\end{bmatrix}
\cdot\begin{bmatrix}
\tilde{M}_{11}&\tilde{M}_{12}\\
\tilde{M}_{21}&\tilde{M}_{22}
\end{bmatrix}\cdot\begin{bmatrix}
X(t_2)\\u(t_2)
\end{bmatrix}\,.
\end{align}

As detailed in \cite{calzetta_hu_2008}, the advantage of working with the $(X,u)$ basis is that here the influence functional takes the compact form 
\begin{align}\label{eq:InfluenceFunctional}
S_{\text{\tiny{IF}}}[X,u](t) =\nonumber\\ 
\int_{t_i}^t dt_1dt_2 &\left[i\frac{ u(t_1)\mathcal{N}^{(\mathcal{L})}(t_1;t_2) u(t_2)}{2}\right.\nonumber\\ 
&\left.+u(t_1)\mathcal{D}^{(\mathcal{L})}(t_1;t_2)X(t_2)\right]\,,
\end{align}
where the noise kernel $\mathcal{N}^{(\mathcal{L})}$ and the dissipation kernel  $\mathcal{D}^{(\mathcal{L})}$ are defined as
\begin{align}
\mathcal{N}^{(\mathcal{L})}(t_1;t_2):=&\frac{g^2}{2\hbar}\left\langle \{\hat{E}(t_1), \hat{E}(t_2)\}\right\rangle_0\,,\nonumber\\
\mathcal{D}^{(\mathcal{L})}(t_1;t_2):=&\frac{ig^2}{\hbar}\left\langle\left[ \hat{E}(t_1), \hat{E}(t_2)\right]\right\rangle_0\theta(t_1-t_2)\,.
\end{align} 
The formal expression \eqref{eq:RedRho_formal} is exact. However, the path integral in the presence of the influence functional is non-trivial to compute. Nevertheless, the equation for the rate of change of the reduced density matrix, i.e., the master equation, can still be obtained perturbatively. The intermediate steps to arrive at the master equation are already detailed in \cite{Gundhi:2023vjs,calzetta_hu_2008}, and here we therefore only write the final result. To leading order, in terms of the noise and the dissipation kernel, the master equation reads
\begin{align}
\label{aeq:MasterEq}
\dot{\hat{\rho}}_r=&-\frac{i}{\hbar}\!\left[\hat{H}^{(\mathcal{L})}_1,\hat{\rho}_{r}(t)\right]\nonumber\\
&-\frac{1}{\hbar}\int_{0}^{t}d\tau \mathcal{N}^{(\mathcal{L})}(t;t-\tau)\left[\hat{s},\left[\hat{s}_{\hat{H}_1}(-\tau),\hat{\rho}_{r}(t)\right]\right]\nonumber\\
&+\frac{i}{2\hbar}\int_{0}^{t}\!
d\tau\mathcal{D}^{(\mathcal{L})}(t;t-\tau)\left[\hat{s},\{\hat{s}_{\hat{H}_1}(-\tau),\hat{\rho}_{r}(t)\}\right]\,,
\end{align} 
where $\hat{s}$, $\hat{H}^{(\mathcal{L})}_1$, $\mathcal{N}^{(\mathcal{L})}$, $\mathcal{D}^{(\mathcal{L})}$ depend upon the Lagrangian used to describe the dynamics. 

\subsection{Master equations for L}
When the dynamics is described by $L$, we have
\begin{align}
\hat{s} = \hat{x}_1\,,\qquad \hat{H}^{L}_1 = \frac{p^2_1}{2m_1}+\frac{g^2x^2_1}{2m_2}\,,\qquad \hat{H}^{L}_2=\frac{p^2_2}{2m_2}\,,
\end{align}
and for the kernels
\begin{align}\label{aeq:KernelsL}
\mathcal{N}^L&= \frac{g^2}{2m_2^2\hbar}\mathrm{Tr}\{\hat{\rho}_2(t_i)\{\hat{p}_2(t_1), \hat{p}_2(t_2)\}\}\,,\nonumber\\
\mathcal{D}^{L}&=\frac{ig^2}{m^2_2\hbar}\mathrm{Tr}\{\hat{\rho}_2(t_i)\left[ \hat{p}_2(t_1), \hat{p}_2(t_2)\right]\}\theta(t_1-t_2)\,.
\end{align}
The Hamiltonians $\hat{H}^L_1$ and $\hat{H}^L_{2}$ govern the free evolution of $\hat{x}(\tau)$ and $\hat{p}_2(\tau)$ respectively, that appear in the master equation~\eqref{aeq:MasterEq}. To second order in $g$, we get 
\begin{align}\label{aeq:KernelsLFact}
&\mathcal{N}^L(t_1;t_2) = \frac{g^2}{m_2^2\hbar}\mathrm{Tr}\{\hat{\rho}_2(t_i)\hat{p}^2_2\}:=\frac{g^2}{m_2^2\hbar}\langle\hat{p}^2_2\rangle_i\nonumber\\
&\mathcal{D}^L= 0\,,\nonumber\\
&\hat{x}_1(-\tau) = \hat{x}_1-\tau\frac{\hat{p}_1}{m_1}.
\end{align}
We point out that $\hat{x}_1(-\tau)$ is only computed to zero$^{\text{th}}$ order in $g$ since it is multiplied with the noise kernel $\mathcal{N}$ and the master equation is only valid upto second order in $g$. Using the relations above in Eq.~\eqref{aeq:MasterEq}, one arrives at the master equation
\begin{align}\label{aeq:Master_L1}
\partial_t\hat{\rho}_r=&-\frac{i}{\hbar}\left[\frac{p^2_1}{2m_1}+\frac{g^2\hat{x}^2_1}{2m_2},\hat{\rho}_r\right]\nonumber\\
&-\frac{g^2\langle\hat{p}^2_2\rangle_i}{\hbar^2m_2^2}\left(t[\hat{x}_1,[\hat{x}_1,\hat{\rho}_r]]{-}\frac{t^2}{2m_1}[\hat{x}_1,[\hat{p}_1,\hat{\rho}_r]]\right).
\end{align} 
For the master equation above, the equation of motion for $\langle\hat{x}_1\rangle_t$ and $\langle\hat{p}_1\rangle_t$ is given by
\begin{align}\label{aeq:EOML1_int}
\frac{d}{dt}\langle\hat{x}_1\rangle_t &= \langle\hat{p}_1\rangle_t/m_1\,,\nonumber\\
\frac{d}{dt}\langle\hat{p}_1\rangle_t &= -g^2\langle\hat{x}_1\rangle_t/m_2\,.
\end{align}
The differential equations above can be easily decoupled and written as
\begin{align}\label{aeq:EOML1}
\frac{d^2}{dt^2}\langle\hat{x}_1\rangle_t+\frac{g^2}{m_1m_2}\langle\hat{x}_1\rangle_t = 0\,,\nonumber\\
\frac{d^2}{dt^2}\langle\hat{p}_1\rangle_t+\frac{g^2}{m_1m_2}\langle\hat{p}_1\rangle_t = 0\,.
\end{align}
\subsection{Master equations for $L'$}
When the dynamics is described by $L'$, we have
\begin{align}
\hat{s} = \hat{p}'_1\,,\qquad \hat{H}^{L'}_1 = \frac{\hat{p}'^2_1}{2m_1}\,,\qquad \hat{H}^{L'}_2=\frac{\hat{p}'^2_2}{2m_2}+\frac{g^2\hat{x}^2_2}{2m_1}\,,
\end{align}
and for the kernels
\begin{align}
\mathcal{N}^{L'}&= \frac{g^2}{2m_1^2\hbar}\mathrm{Tr}\{\hat{\rho}_2(t_i)\{\hat{x}_2(t_1), \hat{x}_2(t_2)\}\}\,,\nonumber\\
\mathcal{D}^{L'}&=\frac{ig^2}{m^2_1\hbar}\mathrm{Tr}\{\hat{\rho}_2(t_i)\left[ \hat{x}_2(t_1), \hat{x}_2(t_2)\right]\}\theta(t_1-t_2)\,.
\end{align} 
Taking into account the free evolution of $\hat{p}'_1$ and $\hat{x}_2$ respectively, upto order second order, we get 
\begin{align}
\mathcal{N}^{L'}(t_1;t_2) ={}&\frac{g^2}{m_1^2\hbar}\mathrm{Tr}\left\lbrace\hat{\rho}_2(t_i)\times\nonumber\right.\\
&\left.\times\left(\hat{x}^2_2+(t_1+t_2)\frac{\{\hat{x}_2,\hat{p}'_2\}}{2m_2}+t_1t_2\frac{\hat{p}'^2_2}{m^2_2}\right)\right\rbrace\nonumber\\
\mathcal{D}^{L'}(t_1;t_2)={}& -\frac{g^2 (t_2-t_1)}{m^2_1m_2}\,,\nonumber\\
\hat{p}'_1(-\tau) ={}&\hat{p}'_1.
\end{align}
Using the relations above in Eq.~\eqref{aeq:MasterEq}, one arrives at the master equation
\begin{align}\label{aeq:Master_L2}
\partial_t{\hat{\rho}'}_1=&-\frac{i}{\hbar}\left[\frac{\hat{p}'^2_1}{2m_1}\left(1-\frac{g^2t^2}{2m_1m_2}\right),\hat{\rho}'_1\right]-\frac{g^2}{\hbar^2m_1^2}\left(t\langle\hat{x}^2_2\rangle_i\phantom{\frac{1}{2}} \right.\nonumber\\
&\left.+\frac{t^3\langle\hat{p}'^2_2\rangle_i}{2m^2_2}+\frac{3t^2}{4m_2}\langle\{\hat{x}_2,\hat{p}'_2\}\rangle_i\right)[\hat{p}'_1,[\hat{p}'_1,\hat{\rho}'_1]].
\end{align}
From the master equation above, the equation of motion for $\langle\hat{x}_1\rangle_t$ and $\langle\hat{p}_1\rangle_t$ is given by
\begin{align}\label{aeq:EOML'1_int}
\frac{d}{dt}\langle\hat{x}_1\rangle_t &= \left(1-\frac{g^2t^2}{2m_1m_2}\right)\langle\hat{p}'_1\rangle_i/m_1\,,\nonumber\\
\frac{d}{dt}\langle\hat{p}'_1\rangle_t &= 0\,.
\end{align}

\section{Equation of motion for $L$, with transformed initial state}\label{appendix:C}
In this section we derive the equation of motion for $\langle\hat{x}_1\rangle_t$, for the Lagrangian $L$, but this time for a non-factorized initial state. As argued in Sec.~\ref{sec:equivalence}, with such a transformation one expects the equation of motion for $\langle\hat{x}_1\rangle_t $ to match the one in Eq.~\eqref{aeq:EOML'1_int} derived from $L'$.

A factorized initial state for the Lagrangian $L'$, 
\begin{align}\label{aeq:RhoIn}
\rho^i_{(L')}(\tilde{x}_{(1,2)},x_{(1,2)})=\rho_{1}(\tilde{x}^i_{1},x^i_{1})\rho_{2}(\tilde{x}^i_{2},x^i_{2})
\end{align}
corresponds to 
\begin{align}\label{aeq:RhoInTrans}
\rho^i_{(L)}&=e^{ig\tilde{x}^i_1\tilde{x}^i_2/\hbar}\rho_{1}(\tilde{x}^i_{1},x^i_{1})\rho_{2}(\tilde{x}^i_{2},x^i_{2})e^{-ig{x}^i_1{x}^i_2/\hbar}\,,\nonumber\\
&=\rho_{1}(\tilde{x}^i_{1},x^i_{1})\Big(e^{ig\tilde{x}^{i}_1\tilde{x}^{i}_2/\hbar}\rho_{2}(\tilde{x}^i_{2},x^i_{2})e^{-ig{x}^i_1{x}^i_2/\hbar}\Big)\,.
\end{align} 
The transformation of the full initial state can be seen as a (system-dependent) phase shift on the environmental initial state alone. The effect of the transformations thus can be fully captured by a modified influence functional due to the modified matrix elements $M_{ab}$. This can be seen from the discussion above Eq.~\eqref{aeq:matrix_elements}: taking the functional derivative with respect to the system variables will produce extra terms at the initial time $t=0$. Therefore, to second order in $g$, we have
\begin{align}
M_{ab}(t_1;t_2)&=\frac{ig^2}{\hbar m^2_2}\int_{\mathrm{tr}}p_{2a}\left(t_1\right) p_{2b}\left(t_2\right) e^{\frac{i}{\hbar}(\tilde{S}_{2}-S_{2})}\rho_{2}^{i}\nonumber\\
&+\frac{ig^2}{\hbar m_2}\int_{\mathrm{tr}}x_{2a}\delta\left(t_1\right) p_{2b}\left(t_2\right) e^{\frac{i}{\hbar}(\tilde{S}_{2}-S_{2})}\rho_{2}^{i}\nonumber\\
&+\frac{ig^2}{\hbar m_2}\int_{\mathrm{tr}}p_{2a}\left(t_1\right) x_{2b}\delta\left(t_2\right) e^{\frac{i}{\hbar}(\tilde{S}_{2}-S_{2})}\rho_{2}^{i}\nonumber\\
&+\frac{ig^2}{\hbar}\int_{\mathrm{tr}}x_{2a}\delta\left(t_1\right) x_{2b}\delta\left(t_2\right) e^{\frac{i}{\hbar}(\tilde{S}_{2}-S_{2})}\rho_{2}^{i}\,.
\end{align} 
Comparing with Eq.~\eqref{aeq:matrix_elements}, where the matrix elements $M_{ab}$ were computed with a factorized initial state, we see that for the Lagrangian $L$, the effects of transformed initial condition can be taken into account by simply replacing $\hat{p}_2(t_1)$ and $\hat{p}_2(t_2)$ with $\hat{p}_2(t_1)\rightarrow \hat{p}_2(t_1)+m_2\hat{x}_2\delta(t_1)$ and $\hat{p}_2(t_2)\rightarrow \hat{p}_2(t_2)+m_2\hat{x}_2\delta(t_2)$ respectively, in the noise and the dissipation kernels appearing in Eq.~\eqref{aeq:KernelsL}~\footnote{In writing $\hat{p}_2(t_1)\rightarrow \hat{p}_2(t_1)+m_2\hat{x}_2\delta(t_1)$, we have adopted the asymmetric normalization $\int_{0}^{\infty}d\tau\delta(\tau) =1$. One can also work with the standard normalization $\int_{0}^{\infty}d\tau\delta(\tau)= 1/2$ by writing $\hat{p}_2(t_1)\rightarrow \hat{p}_2(t_1)+2m_2\hat{x}_2\delta(t_1)$.}. Since our goal is to study the time evolution of $\langle \hat{x}_1 \rangle $ and $\langle \hat{p}_1 \rangle$, we only need to look at the transformation of the dissipation kernel $\mathcal{D}^L$, as $\mathcal{N}^L$ does not affect the time evolution of $\langle \hat{x}_1 \rangle $ or $\langle \hat{p}_1 \rangle $. Under the transformed, non-factorized initial state (Eq.~\eqref{aeq:RhoInTrans}), for the conjugate variables corresponding to $L$, we get that the dissipation kernel goes from being zero in Eq.~\eqref{aeq:KernelsLFact} to
\begin{align}
\mathcal{D}^L(t_1;t_2) = \frac{g^2}{m_2}(\delta(t_2)-\delta(t_1))\,.
\end{align}
The non-zero dissipation kernel adds an additional term to the equation of motion for $\langle\hat{p}_1\rangle_t$, such that 
\begin{align}
&\mathrm{Tr}\{\dot{\hat{\rho}}_r\hat{p}_1\} = -g^2\langle\hat{x}_1\rangle_t/m_2+\nonumber\\
&\frac{ig^2}{2m_2\hbar}\int_{0}^{t}\!
d\tau(\delta(t-\tau)-\delta(t))\mathrm{Tr}\{\hat{p}_1\left[\hat{x}_1,\{\hat{x}_1(-\tau),\hat{\rho}_{r}(t)\}\right]\}\,.
\end{align}
The $\delta(t)$ term in the equation above does not contribute, since after integrating over $\tau$ one would have terms involving $t\delta(t)$ and $t^2\delta(t)$. Thus, to order $g^2$, the equation of motion for $\hat{p}_1$, for the Lagrangian $L$ simplifies to
\begin{align}
\frac{d}{dt}\langle\hat{p}_1\rangle_t= &-\frac{g^2\langle\hat{x}_1\rangle_t}{m_2}\nonumber\\
&+\frac{g^2}{2m_2}\int_{0}^{t}\!
d\tau\delta(t-\tau)\mathrm{Tr}\left\lbrace\hat{x}_1-\frac{\tau \hat{p}_1}{m_1},\hat{\rho}_{r}(t)\right\rbrace\nonumber\\
=&-\frac{g^2\langle\hat{x}_1\rangle_t}{m_2}+\frac{g^2\langle\hat{x}_1\rangle_t}{m_2}-\frac{g^2t\langle\hat{p}_1\rangle_t}{m_1m_2}
\end{align}
To second order in $g$, the solution for $\langle\hat{p}_1\rangle_t$ reads
\begin{align}
\langle\hat{p}_1\rangle_t=\left(1-\frac{g^2t^2}{2m_1m_2}\right)\langle\hat{p}_1\rangle_i\,.
\end{align}
Thus, the equation of motion for the conjugate variables of particle 1 for the Lagrangian, with non-factorized transformed initial conditions read
\begin{align}\label{aeq:EOML_Trans}
\frac{d}{dt}\langle\hat{x}_1\rangle_t &= \left(1-\frac{g^2t^2}{2m_1m_2}\right)\langle\hat{p}_1\rangle_i/m_1\,,\nonumber\\
\langle\hat{p}_1\rangle_t&=\left(1-\frac{g^2t^2}{2m_1m_2}\right)\langle\hat{p}_1\rangle_i\,.
\end{align}
We see that the equation on motion for $\langle \hat x_1\rangle_t$ is similar to Eq.~\eqref{aeq:EOML'1_int}, except for the presence of $\langle\hat{p}_1\rangle_i$ instead of $\langle\hat{p}'_1\rangle_i$.
While the two operators $\hat p_1$ and $\hat p'_1$ are different, within their own representation they both act as $-i\hbar\frac{\partial}{\partial x_1}$. Therefore, since the two averages are computed on the same initial density matrix for particle 1 (see Eq.~\eqref{aeq:RhoIn} and Eq.~\eqref{aeq:RhoInTrans}) we have that, numerically, $\langle\hat{p}_1\rangle_i = \langle\hat{p}'_1\rangle_i$. We underline that this is true at the initial time only. Thus the equation of motion for $\langle x_1\rangle_t$ obtained from $L$ with a transformed initial state is exactly the same as that given in the primed representation~\eqref{aeq:EOML'1_int}, as anticipated.

\section{Master equation and variances for bremsstrahlung}\label{appendix:D}
The master equation (ignoring free evolution) is given by
\begin{align}\label{SEQ:MasterEquation}
\partial_t\hat{\rho}_r=&-\frac{1}{\hbar}\int_{0}^{t} d\tau \mathcal{N}_{0}(\tau)\left[\hat{x},\left[\hat{x}_{\text{\tiny{H}}_s}(-\tau),\hat{\rho}_{r}(t)\right]\right]\nonumber\\
&+\frac{i}{2\hbar}\int_{0}^{t}
d\tau\mathcal{D}(\tau)\left[\hat{x},\{\hat{x}_{\text{\tiny{H}}_s}(-\tau),\hat{\rho}_{r}(t)\}\right]\,.
\end{align} 
We recall the relations
\begin{align}
\hat{x}_{\text{\tiny{H}}_s}(-\tau)=\cos(\Omega\tau)\hat{x}-\sin(\Omega\tau)\frac{\hat{p}}{m\Omega}\,,
\end{align}
and
\begin{align}
\mathcal{N}_{0}(\tau)&= \frac{4\alpha\hbar}{\pi c^2}\frac{\left(\epsilon^4-6\epsilon^2\tau^2+\tau^4\right)}{\left(\epsilon^2+\tau^2\right)^4}\,,\nonumber\\
\mathcal{D}(\tau)&=\frac{4\hbar\alpha}{3c^2}\theta(\tau)\delta'''_{\epsilon}(\tau)\,.
\end{align} 
On time scales $t\gg\epsilon$ the function $\delta_{\epsilon}(\tau)$ can be effectively treated as a Dirac delta. The integral $\mathcal{D}(\tau)$ can be evaluated by integrating by parts. In shifting the derivatives from $\delta_{\epsilon}$ onto $\hat{x}_{\text{\tiny{H}}_s}(-\tau)$, we get two non-zero boundary terms. The one proportional to $\delta''_{\epsilon}(0)$ cancels the contribution coming from $\hat{V}_{\text{\tiny{EM}}}$ in $\hat{H}_s$ while the one proportional to $\delta_{\epsilon}(0)$ leads to the renormalization of mass/frequency. Performing the integration involving the noise and the dissipation kernels, we get

\begin{align}\label{SEQ:MasterEqSimp}
&\partial_{t}\hat{\rho}_r = -\frac{i}{\hbar}\left[\frac{\hat{p}^2}{2m}+\frac{1}{2}m\Omega^2_{r}\hat{x}^2,\hat{\rho}_{r}\right]\nonumber\\
& -\frac{i\alpha\Omega^2}{3mc^2}\left[\hat{x},\{\hat{p},\hat{\rho}_r\}\right]
-\frac{\alpha\Omega^3}{3c^2}[\hat{x},[\hat{x},\hat{\rho}_r]] \nonumber\\
&-\frac{2\alpha}{mc^2}\left(\frac{1}{3\pi \epsilon^2}-\frac{\Omega^2}{3\pi}\log\left(\epsilon \Omega\right)- \frac{\gamma_\text{\tiny{E}}\Omega^2}{3\pi}\right)[\hat{x},[\hat{p},\hat{\rho}_r]]\,,
\end{align}
where $\Omega^2_{r}:=\Omega^2\left(1-4\alpha\hbar\omega_{\mathrm{max}}/(3\pi m c^2)\right)$. Note that the distinction between $\Omega$ and $\Omega_r$ would only give a higher order contribution to the master equation in terms other than the free evolution, and is therefore not relevant since the master equation in our work is derived upto $e^2$. Therefore, wherever $\Omega$ appears in the dynamics, it can be treated as the renormalized frequency.

We now proceed with the calculation of the evolution of the variances of the system. As mentioned in the main text, we will neglect the term that scales explicitly with the cutoff, as it is part of the dressing, and we regularize the logarithmic divergence by setting $\epsilon \approx \frac{\hbar}{mc^2}$.
The coupled equations governing their dynamics are derived from Eq.~\eqref{SEQ:MasterEqSimp} to be
\begin{align}
&\frac{d}{dt}\langle\hat{x}^2\rangle = \frac{1}{m}\langle\lbrace\hat{x},\hat{p}\rbrace\rangle\,,\nonumber\\
&\frac{d}{dt}\langle\hat{p}^2\rangle = -m\Omega^2\langle\lbrace\hat{x},\hat{p}\rbrace\rangle-\frac{4\alpha\hbar\Omega^2}{3mc^2}\langle\hat{p}^2\rangle+\frac{2\alpha\hbar^2 \Omega^3}{3c^2}\,,\nonumber\\
&\frac{d}{dt}\langle\lbrace\hat{x},\hat{p}\rbrace\rangle =\frac{2}{m}\langle\hat{p}^2\rangle-2m\Omega^2\langle\hat{x}^2\rangle -\frac{2\alpha\hbar\Omega^2}{3mc^2}\langle{\lbrace\hat{x},\hat{p}\rbrace}\rangle\nonumber\\
&+\frac{4\alpha\gamma_\text{\tiny{E}}\hbar^2\Omega^2}{3\pi mc^2} + \frac{4\alpha\hbar^2\Omega^2}{3\pi m  c^2}\log\left(\frac{\hbar\Omega}{mc^2}\right)\,.
\end{align}
To obtain the stationary variances, we simply put all the derivatives on the left hand side to zero, to obtain
\begin{align}
&\frac{\langle\hat{p}^2\rangle}{2m} = \frac{\hbar\Omega}{4}\,,\\ 
&\frac{1}{2}m\Omega^2\langle\hat{x}^2\rangle = \frac{\langle\hat{p}^2\rangle}{2m}+\frac{\alpha\hbar\Omega}{3\pi}\left(\frac{\hbar\Omega}{mc^2}\right)\left(\gamma_\text{\tiny{E}} + \log\left(\frac{\hbar\Omega}{mc^2}\right)\right)\,.\label{eq:variance}
\end{align}
We see that the last two terms on the right hand side of Eq.~\eqref{eq:variance} would only give a significant contribution when $\hbar\Omega\simeq mc^2$. However, since the calculations are performed within the non-relativistic regime, we must have $\hbar\Omega\ll mc^2$ for consistency. These terms originate exclusively from the last line in Eq.~\eqref{SEQ:MasterEqSimp} and it is in this sense that, for all practical purposes, the term involving $[\hat{x},[\hat{p},\hat{\rho}_r]]$ in the master equation can be ignored. This is because it neither contributes at the level of the expectation values of $\hat{x}$ and $\hat{p}$, nor at the level of their variances. The equations above are also consistent with the equipartition theorem.

\bibliography{OQS_Lagrangians_accepted}

\end{document}